\errorstopmode
\input amssym.def
\input amssym.tex

% Page layout

\magnification=\magstephalf
\hsize=14.0 true cm
\vsize=19 true cm
\hoffset=1.0 true cm
\voffset=2.0 true cm

\abovedisplayskip=12pt plus 3pt minus 3pt
\belowdisplayskip=12pt plus 3pt minus 3pt
\parindent=1.0em

% Fonts

\font\sixrm=cmr6
\font\eightrm=cmr8
\font\ninerm=cmr9

\font\sixi=cmmi6
\font\eighti=cmmi8
\font\ninei=cmmi9

\font\sixsy=cmsy6
\font\eightsy=cmsy8
\font\ninesy=cmsy9

\font\sixbf=cmbx6
\font\eightbf=cmbx8
\font\ninebf=cmbx9

\font\eightit=cmti8
\font\nineit=cmti9

\font\eightsl=cmsl8
\font\ninesl=cmsl9

\font\sixss=cmss8 at 8 true pt
\font\sevenss=cmss9 at 9 true pt
\font\eightss=cmss8
\font\niness=cmss9
\font\tenss=cmss10

\font\sixmib=cmmib6
\font\sevenmib=cmmib7
\font\eightmib=cmmib8
\font\ninemib=cmmib9
\font\tenmib=cmmib10

 at 12 true pt
 at 12 true pt
\font\bigrm=cmr10 at 12 true pt
 at 12 true pt
 at 12 true pt

 at 16 true pt
%\font\Bigsy=cmsy12 at 16 true pt
%\font\Bigex=cmex12 at 16 true pt
 at 16 true pt
\font\Bigrm=cmr12 at 16 true pt
 at 16 true pt
 at 16 true pt

\catcode`@=11
\newfam\ssfam
\newfam\mibfam

\def\tenpoint{\def\rm{\fam0\tenrm}%
    \textfont0=\tenrm \scriptfont0=\sevenrm \scriptscriptfont0=\fiverm
    \textfont1=\teni  \scriptfont1=\seveni  \scriptscriptfont1=\fivei
    \textfont2=\tensy \scriptfont2=\sevensy \scriptscriptfont2=\fivesy
    \textfont3=\tenex \scriptfont3=\tenex   \scriptscriptfont3=\tenex
    \textfont\itfam=\tenit                  \def\it{\fam\itfam\tenit}%
    \textfont\slfam=\tensl                  \def\sl{\fam\slfam\tensl}%
    \textfont\bffam=\tenbf \scriptfont\bffam=\sevenbf
                           \scriptscriptfont\bffam=\fivebf
                           \def\bf{\fam\bffam\tenbf}%
    \textfont\ssfam=\tenss \scriptfont\ssfam=\sevenss
                           \scriptscriptfont\ssfam=\sevenss
                           \def\ss{\fam\ssfam\tenss}%
    \textfont\mibfam=\tenmib \scriptfont\mibfam=\sevenmib
                             \scriptscriptfont\mibfam=\sevenmib
                             \def\mib{\fam\mibfam\tenmib}%
    \normalbaselineskip=13pt
    \setbox\strutbox=\hbox{\vrule height8.5pt depth3.5pt width0pt}%
    \let\big=\tenbig
    \normalbaselines\rm}

\def\ninepoint{\def\rm{\fam0\ninerm}%
    \textfont0=\ninerm      \scriptfont0=\sixrm
                            \scriptscriptfont0=\fiverm
    \textfont1=\ninei       \scriptfont1=\sixi
                            \scriptscriptfont1=\fivei
    \textfont2=\ninesy      \scriptfont2=\sixsy
                            \scriptscriptfont2=\fivesy
    \textfont3=\tenex       \scriptfont3=\tenex
                            \scriptscriptfont3=\tenex
    \textfont\itfam=\nineit \def\it{\fam\itfam\nineit}%
    \textfont\slfam=\ninesl \def\sl{\fam\slfam\ninesl}%
    \textfont\bffam=\ninebf \scriptfont\bffam=\sixbf
                            \scriptscriptfont\bffam=\fivebf
                            \def\bf{\fam\bffam\ninebf}%
    \textfont\ssfam=\niness \scriptfont\ssfam=\sixss
                            \scriptscriptfont\ssfam=\sixss
                            \def\ss{\fam\ssfam\niness}%
    \textfont\mibfam=\ninemib \scriptfont\mibfam=\sixmib
                            \scriptscriptfont\mibfam=\sixmib
                            \def\mib{\fam\mibfam\ninemib}%
    \normalbaselineskip=12pt
    \setbox\strutbox=\hbox{\vrule height8.0pt depth3.0pt width0pt}%
    \let\big=\ninebig
    \normalbaselines\rm}

\def\eightpoint{\def\rm{\fam0\eightrm}%
    \textfont0=\eightrm      \scriptfont0=\sixrm
                             \scriptscriptfont0=\fiverm
    \textfont1=\eighti       \scriptfont1=\sixi
                             \scriptscriptfont1=\fivei
    \textfont2=\eightsy      \scriptfont2=\sixsy
                             \scriptscriptfont2=\fivesy
    \textfont3=\tenex        \scriptfont3=\tenex
                             \scriptscriptfont3=\tenex
    \textfont\itfam=\eightit \def\it{\fam\itfam\eightit}%
    \textfont\slfam=\eightsl \def\sl{\fam\slfam\eightsl}%
    \textfont\bffam=\eightbf \scriptfont\bffam=\sixbf
                             \scriptscriptfont\bffam=\fivebf
                             \def\bf{\fam\bffam\eightbf}%
    \textfont\ssfam=\eightss \scriptfont\ssfam=\sixss
                             \scriptscriptfont\ssfam=\sixss
                             \def\ss{\fam\ssfam\eightss}%
    \textfont\mibfam=\eightmib \scriptfont\mibfam=\sixmib
                             \scriptscriptfont\mibfam=\sixmib
                             \def\mib{\fam\mibfam\eightmib}%
    \normalbaselineskip=10pt
    \setbox\strutbox=\hbox{\vrule height7.0pt depth2.0pt width0pt}%
    \let\big=\eightbig
    \normalbaselines\rm}

\def\tenbig#1{{\hbox{$\left#1\vbox to8.5pt{}\right.\n@space$}}}
\def\ninebig#1{{\hbox{$\textfont0=\tenrm\textfont2=\tensy
                       \left#1\vbox to7.25pt{}\right.\n@space$}}}
\def\eightbig#1{{\hbox{$\textfont0=\ninerm\textfont2=\ninesy
                       \left#1\vbox to6.5pt{}\right.\n@space$}}}

\font\sectionfont=cmbx10
\font\subsectionfont=cmti10

\def\figurecaptionfont{\ninepoint}
\def\tablecaptionfont{\ninepoint}
\def\footnotefont{\eightpoint}

% New count registers

\newcount\equationno
\newcount\bibitemno
\newcount\figureno
\newcount\tableno

\equationno=0
\bibitemno=0
\figureno=0
\tableno=0
%\advance\pageno by -1

% Footline

\footline={\ifnum\pageno=0{\hfil}\else
{\hss\rm\the\pageno\hss}\fi}

% Section macro

\def\section #1. #2 \par
{\vskip0pt plus .10\vsize\penalty-100 \vskip0pt plus-.10\vsize
\vskip 1.6 true cm plus 0.2 true cm minus 0.2 true cm
\global\def\equationlabel{#1}
\global\equationno=0
\leftline{\sectionfont #1. #2}\par
\immediate\write\terminal{Section #1. #2}
\vskip 0.7 true cm plus 0.1 true cm minus 0.1 true cm
\noindent}

% Subsection macro

\def\subsection #1 \par
{\vskip0pt plus 0.8 true cm\penalty-50 \vskip0pt plus-0.8 true cm
\vskip2.5ex plus 0.1ex minus 0.1ex
\leftline{\subsectionfont #1}\par
\immediate\write\terminal{Subsection #1}
\vskip1.0ex plus 0.1ex minus 0.1ex
\noindent}

% Appendix macro

\def\appendix #1. #2 \par
{\vskip0pt plus .10\vsize\penalty-100 \vskip0pt plus-.10\vsize
\vskip 1.6 true cm plus 0.2 true cm minus 0.2 true cm
\global\def\equationlabel{\hbox{\rm#1}}
\global\equationno=0
\leftline{\sectionfont Appendix #1. #2}\par
\immediate\write\terminal{Appendix #1. #2}
\vskip 0.7 true cm plus 0.1 true cm minus 0.1 true cm
\noindent}

%\def\appendix #1. #2 \par
%{\vskip0pt plus .20\vsize\penalty-100 \vskip0pt plus-.20\vsize
%\vskip 1.6 true cm plus 0.2 true cm minus 0.2 true cm
%\global\def\equationlabel{\hbox{\rm#1}}
%\global\equationno=0
%\leftline{\sectionfont Appendix #1. #2}\par
%\immediate\write\terminal{Appendix #1. #2}
%\vskip 0.7 true cm plus 0.1 true cm minus 0.1 true cm
%\noindent}

% Displayed equations

\def\equation#1{$$\displaylines{\qquad #1}$$}
\def\enum{\global\advance\equationno by 1
\hfill\llap{{\rm(\equationlabel.\the\equationno)}}}

% Bibliography macro, references

\def\ifundefined#1{\expandafter\ifx\csname#1\endcsname\relax}

\def\ref#1{\ifundefined{#1}?\immediate\write\terminal{unknown reference
on page \the\pageno}\else\csname#1\endcsname\fi}

\newwrite\terminal
\newwrite\bibitemlist

\def\bibitem#1#2\par{\global\advance\bibitemno by 1
\immediate\write\bibitemlist{\string\def
\expandafter\string\csname#1\endcsname
{\the\bibitemno}}
\item{[\the\bibitemno]}#2\par}

\def\beginbibliography{
\vskip0pt plus .15\vsize\penalty-100 \vskip0pt plus-.15\vsize
\vskip 1.2 true cm plus 0.2 true cm minus 0.2 true cm
\leftline{\sectionfont References}\par
\immediate\write\terminal{References}
\immediate\openout\bibitemlist=biblist
\frenchspacing\parindent=1.8em
\vskip 0.5 true cm plus 0.1 true cm minus 0.1 true cm}

\def\endbibliography{
\immediate\closeout\bibitemlist
\nonfrenchspacing\parindent=1.0em}

% Figure and table captions

\def\figurecaption#1{\global\advance\figureno by 1
\narrower\figurecaptionfont
Fig.~\the\figureno. #1}

\def\tablecaption#1{\global\advance\tableno by 1
\vbox to 0.25 true cm { }
\centerline{\tablecaptionfont%
Table~\the\tableno. #1}
\vskip-0.4 true cm}

\def\thicktablerule{\hrule height0.8pt}
\def\thintablerule{\hrule height0.4pt}

\tenpoint

\immediate\openin\bibitemlist=biblist
\ifeof\bibitemlist\immediate\closein\bibitemlist
\else\immediate\closein\bibitemlist
\input biblist \fi

% current year and month

\def\thismonth{\ifcase\month\or
January\or February\or March\or April\or May\or June\or
July\or August\or September\or October\or November\or December\fi}

\input epsf
\epsfclipon

% Definitions and abbreviations

% Roman letters in math formulae

% Real and integer numbers

% Special relations and symbols

\def\proof{\noindent{\sl Proof:}\kern0.6em}

\def\frac#1#2{\hbox{$#1\over#2$}}
\def\dual{\mathstrut^*\kern-0.1em}

\def\lvec#1{\setbox0=\hbox{$#1$}
    \setbox1=\hbox{$\scriptstyle\leftarrow$}
    #1\kern-\wd0\smash{
    \raise\ht0\hbox{$\raise1pt\hbox{$\scriptstyle\leftarrow$}$}}
    \kern-\wd1\kern\wd0}
\def\rvec#1{\setbox0=\hbox{$#1$}
    \setbox1=\hbox{$\scriptstyle\rightarrow$}
    #1\kern-\wd0\smash{
    \raise\ht0\hbox{$\raise1pt\hbox{$\scriptstyle\rightarrow$}$}}
    \kern-\wd1\kern\wd0}
\def\slash#1{\setbox0=\hbox{$#1$}\setbox1=\hbox{$\kern1pt/$}
    #1\kern-\wd0\kern1pt/\kern-\wd1\kern\wd0}

% Lattice derivatives

\def\nabstar#1{{\nabla\kern0.5pt\smash{\raise 4.5pt\hbox{$\ast$}}
               \kern-5.5pt_{#1}}}

\def\drvstar#1{{\partial\kern0.5pt\smash{\raise 4.5pt\hbox{$\ast$}}
               \kern-6.0pt_{#1}}}

\def\ldrvstar#1{{\lvec{\,\partial}\kern-0.5pt\smash{\raise 4.5pt\hbox{$\ast$}}
               \kern-5.0pt_{#1}}}

% Units

% Constants

% Fields

% Dirac matrices

\def\diracstar#1#2{
    \setbox0=\hbox{$\gamma$}\setbox1=\hbox{$\gamma_{#1}$}
    \gamma_{#1}\kern-\wd1\kern\wd0
    \smash{\raise4.5pt\hbox{$\scriptstyle#2$}}}

% Gauge group

\def\Ad{{\rm Ad}\kern0.1em}

% Masses and decay constants

\def\mpi{M_{\pi}}
\def\Mpi{\mpi}
\def\msea{m_{\rm sea}}

\def\ksea{\kappa_{\rm sea}}

% Parameters and abbreviations

\def\csw{c_{\rm sw}}

\def\Ngcr{N_{\rm GCR}}
\def\Nm{N_{s}}
\def\epstwo{\varepsilon\hbox{\vrule height7.0pt depth1.5pt width0pt}_2}

%
%\vbox{\vskip0.0cm}
\rightline{CERN-PH-TH/2007-200}

\vskip1.5cm 
\centerline{\Bigrm
Deflation acceleration of lattice QCD simulations
}
%\vskip0.2cm
%\centerline{\Bigrm
%at small quark masses
%}
%
\vskip 0.6 true cm
\centerline{\bigrm Martin L\"uscher}
\vskip1ex
\centerline{\it CERN, Physics Department, TH Division}
\centerline{\it CH-1211 Geneva 23, Switzerland}
\vskip 0.8 true cm
\thintablerule
\vskip 2.0ex
\ninepoint
\leftline{\bf Abstract}
\vskip 1.0ex\noindent
Close to the chiral limit, many calculations in numerical lattice QCD
can potentially be accelerated using low-mode deflation techniques.
In this paper it is shown that the recently introduced 
domain-decomposed deflation subspaces can be propagated along
the field trajectories generated by the Hybrid Monte Carlo (HMC) 
algorithm with a modest effort.
The quark forces that drive the simulation may then be computed 
using a deflation-accelerated solver for the lattice Dirac equation.
As a consequence, the computer time required for the simulations
is significantly reduced and an improved scaling behaviour
of the simulation algorithm with respect to the quark mass 
is achieved.
\vskip 2.0ex
\thintablerule

\tenpoint

\vskip-0.3cm

\section 1. Introduction

Numerical simulations of lattice QCD have become
a powerful tool for quantitative studies of 
the properties of the strongly interacting particles.
The systematic errors in these calculations
are not easy to assess, however, and most results 
published to date must for this reason be regarded as
preliminary.
In order to remove this deficit,
simulations in a wide range of lattice
spacings, lattice volumes and sea-quark masses will have to be performed,
thus requiring larger and larger lattices to be considered.

The use of advanced simulation techniques that scale well with
the quark masses and the lattice parameters
is likely to be crucial for the success of this programme. 
In the last few years,
a lot of progress has already been made in this direction.
In particular, preconditioned [\ref{Hasenbusch}--\ref{UrbachEtAl}]
and other variants [\ref{RHMC}]
of the Hybrid Monte Carlo (HMC) algorithm [\ref{HMC}] were introduced 
which allowed simulations to be performed at much smaller 
quark masses than was possible before (see
refs.~[\ref{QCDlite}--\ref{Ukita}], for example).

Important advances have also been made in the area of 
variance-reduction methods and quark propagator calculations.
The progress here is often
based on deflation ideas, where
the low modes of the lattice Dirac operator are 
treated separately from the bulk of the modes
[\ref{deForcrand}--\ref{DDDeflation}].
Usually the mode separation is achieved
by calculating the low-lying eigenvalues and the 
associated eigenvectors of the Dirac operator. 
When implemented in this way, low-mode deflation 
however tends to become inefficient or even impractical
on large lattices,
because the number of eigenvectors that must be computed
grows proportionally to the lattice volume.

This limitation can be overcome through the use of
domain-decomposed deflation subspaces and projection
techniques that do not require the deflation subspace to be
spanned by eigenvectors of the Dirac operator [\ref{DDDeflation}].
Domain-decomposed deflation subspaces 
can be generated with a modest computational effort and
were found to be highly effective. The numerical solution 
of the lattice Dirac equation, for example,
can be accelerated by an impressive factor using such deflation 
subspaces together with a suitable preconditioner [\ref{DDDeflation}].

In the present paper it will be shown that domain-decomposed deflation
subspaces can be easily propagated along the molecular-dynamics 
trajectories generated by the HMC algorithm. At each step of 
the molecular-dynamics evolution, the subspace may then be used
to accelerate the computation of the quark forces that drive
the simulation. The simulation
algorithm itself thus remains the same,
but the required computer time
is reduced significantly, particularly so at small quark masses.
Moreover, a further acceleration can be achieved through the simultaneous
use of the ``chronological inversion method'' 
of Brower et al.~[\ref{BrowerEtAl}].

\section 2. Preliminaries

The deflation acceleration of the HMC algorithm will be worked
out for a definite choice of the lattice formulation of the theory and 
of the preconditioned form of the algorithm.
However, low-mode deflation and the subspace propagation
do not depend on the particular choices made and are expected to be 
widely applicable.

\subsection 2.1 Lattice formulation

The standard O($a$)-improved Wilson theory [\ref{SW},\ref{OaImp}]
will be considered, with a doublet of
mass-degenerate sea quarks and non-perturbatively determined coefficient $\csw$
of the Sheikholeslami--Wohlert improvement term [\ref{NPimp}].
In this theory, the adjustable parameters are the lattice sizes
$T$ and $L$ in the time and the space directions,
the inverse bare coupling
$\beta$ and the bare sea-quark mass in units of the lattice spacing 
(or, equivalently, the hopping parameter $\ksea$).

The utility of any particular deflation method
in general depends on the physical situation one is interested in.
Here it will be assumed that the theory is considered 
in the large-volume regime
of QCD, where, say, $T\geq L\geq2$ fm and $\mpi L\geq3$, 
$\mpi$ being the mass of the ``pion'' at the specified sea-quark mass.
Moreover, in order to guarantee the stability
of the HMC simulations,
the lattice parameters must be such
that the low end of the spectrum of the Dirac operator is 
safely separated from the origin [\ref{Stability}].

\subsection 2.2 DD-HMC algorithm

The HMC algorithm is 
nowadays practically
only used in preconditioned form, where the 
quark determinant is split into several factors before
the pseudo-fermion fields are introduced
[\ref{Hasenbusch}--\ref{RHMC}].
The associated forces must then be
computed at regular intervals along the generated 
trajectories in field space, the larger forces more often than the 
smaller ones, according to a scheme introduced
by Sexton and Weingarten [\ref{SextonWeingarten}].

In the case of the DD-HMC algorithm (which is the algorithm 
chosen here)\kern1.5pt\footnote{$\dagger$}{\footnotefont%
The acronym DD-HMC stands for ``Domain Decomposition Hybrid Monte Carlo'',
which is the now commonly adopted name of the Schwarz-preconditioned  
HMC algorithm that was introduced in ref.~[\ref{SchwarzIII}].},
the lattice is decomposed into non-overlapping blocks of lattice points
and the quark determinant is factorised into the product of the 
determinants of the block Dirac operators times another factor
that couples the gauge fields on the different blocks. There are
thus two kinds of quark forces, the block forces and the 
block-interaction force, the latter being the one that includes
the contributions of the low modes of the Dirac operator.

The computation of the block-interaction force requires
the Dirac equation on the full lattice to be solved for 
two source fields.
This calculation usually consumes
the dominant fraction of the computer time spent for the simulation.
The equation will here be solved using the 
Schwarz-preconditioned GCR algorithm [\ref{SchwarzII}], or its
deflated version [\ref{DDDeflation}] once the deflation subspace
is available along the molecular-dynamics trajectories.

\subsection 2.3 Domain-decomposed deflation subspaces

For the reader's convenience and in order to set up the notation,
the definition of domain-decomposed deflation
subspaces is recalled in the following paragraphs
and it is explained how they are obtained in practice [\ref{DDDeflation}].

The starting point is again a regular decomposition of the lattice into
non-overlap\-ping 
blocks $\Lambda$ of fixed size 
(equal to $4^4$ or $6^2\times4^2$, for example).
Similar block de\-com\-positions of the lattice
are also used for the preconditioning of the HMC algorithm
and the GCR solver, but all these block grids are logically 
unrelated and it is better to think of them as being separate
structures even if the block sizes happen to be the same.

Once a particular block decomposition is selected,
a set $\psi_l(x)$, $l=1,\ldots,\Nm$, of quark fields is
generated through the smoothing procedure outlined below. The
fields are then projected
to the blocks $\Lambda$ through 
\equation{
  \psi^{\Lambda}_l(x)=\cases{\psi_l(x)& if $x\in\Lambda$,\cr
                             \noalign{\vskip1ex}
                                  0   & otherwise,\cr}
  \enum
}
and the linear space spanned by the set of all block 
fields $\psi^{\Lambda}_l(x)$ is
taken to be the deflation subspace.
The latter thus has
dimension equal to $\Nm$ times the number of blocks.
Typical values of $\Nm$ range from 12 to 24,
but as explained in ref.~[\ref{DDDeflation}],
some tuning is normally required to find the 
best values of $\Nm$.

The smoothing procedure starts from some random
fields $\psi_l(x)$, $l=1,\ldots,\Nm$, and consists in applying
a number of approximate inverse iteration steps 
\equation{
   \psi_{l}(x)\to\hbox{``}D^{-1}\kern1pt\hbox{''}\psi_{l}(x)
   \enum
}
to them, where $D$ denotes the lattice Dirac operator at a value of 
the bare quark
mass close to the critical mass
(the inverse of $D$ is put in
quotes in this formula in order to make it clear that an approximation to the 
inverse is being applied). As a result the components of the fields 
along the high modes of the Dirac operator are suppressed and 
the fields will therefore have a strong overlap with the subspace 
spanned by low modes of the operator. The local coherence of the latter
then guarantees that the associated domain-decomposed deflation subspace
is highly effective [\ref{DDDeflation}].

Exactly which algorithm is used for the approximate solution
of the Dirac equation in the recursion (2.2) should not matter 
too much. The procedure employed here is 
described in appendix A.

\section 3. Acceleration of the DD-HMC algorithm

In the following, the basic strategy will be to 
generate a domain-decomposed deflation subspace
at the beginning of each molecular-dynamics trajectory
and to preserve its deflation efficiency along the trajectory 
by applying a suitable update procedure.
The subspace is then used to speed up the computation 
of the block-interaction force and thus to accelerate
the simulation. This section also includes a short
description of the chronological inversion method of Brower 
et al.~[\ref{BrowerEtAl}], which is recommended to be used 
together with the deflation acceleration.

\subsection 3.1 Subspace update procedure

The effectiveness of any particular update procedure 
can be determined by observing the number $\Ngcr$ of 
deflated solver iterations that are
required for the computation of the block-interaction force.
The typical behaviour of the iteration number
along a mo\-le\-cular-dynamics trajectory is shown in fig.~1.
One actually observes two iteration numbers in such an experiment,
represented by the lower and upper series of points in the figure,
since the Dirac equation must be solved two times when the 
force is calculated.

\topinsert
\vbox{
\vskip0.0cm
\epsfxsize=11.5cm\hskip0.25cm\epsfbox{plots/GCRhist.eps}
\vskip0.3cm
\figurecaption{%
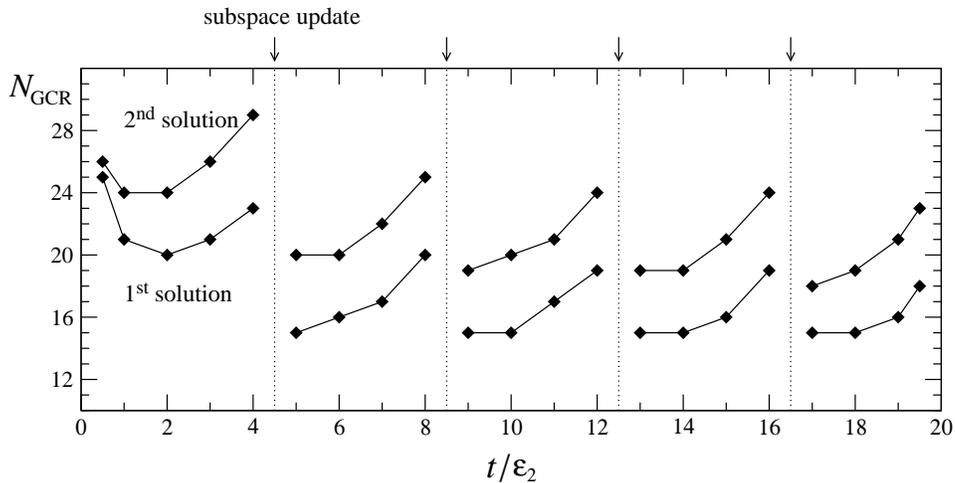
Typical history of the
iteration numbers $\Ngcr$ (data points) 
of the deflated Schwarz-pre\-con\-ditioned GCR solver
along a molecular-dynamics 
trajectory on a $64\times32^3$ lattice at $\ksea=0.13625$,
plotted against the molecular-dynamics time $t$ given in units of the 
integration step size $\epstwo$
(the lattice and algorithmic
parameters are fully specified in sect.~4).
}
\vskip0.3cm
}
\endinsert

In the example shown here, each update of the deflation 
subspace consists in applying
a single deflated inverse iteration step to the quark 
fields $\psi_l(x)$, $l=1,\ldots,\Nm$, that define the subspace 
via the block projection (2.1)
(deflated inverse iteration is explained in detail in appendix A).
After every subspace update, the solver iteration numbers drop to
a lower value and then slowly increase again as the gauge field
evolves along the trajectory. The behaviour is different in
the initial period of the molecular-dynamics integration
for a reason explained in the next subsection.

As it turns out, this simple procedure works well at all values of the 
lattice spacing and the sea-quark mass considered so far.
The optimal number of subspace updates varies, but
in order to maintain a high deflation efficiency of the subspace
it proves to be entirely sufficient to update the fields $\psi_l(x)$ through
the application of a single deflated inverse iteration step.

The frequency of the subspace updates along the
molecular-dynamics trajectories could be taken to be a static
parameter whose value is chosen based on experience.
However, it is more elegant and probably also
more efficient to let the simulation algorithm
choose when the subspace is to be updated.
Such an automatic update procedure is described in appendix B.

\subsection 3.2 Chronological inversion method

The solutions of the Dirac equation calculated in the course
of a molecular-dynamics trajectory evolve smoothly with time
and one can try to forecast the solution at the next integration
step from the previous solutions.
The forecast will normally not be as accurate as required, 
but the solver can start from the proposed solution and will
obtain the solution to the specified precision faster than when
it starts from zero.

The number of previous solutions that should be kept in memory
is an adjustable parameter whose optimal value depends on 
the lattice and algorithmic parameters as well as on
which Dirac equation is being considered. In the case of the 
DD-HMC algorithm, there are altogether three equations, the normal 
even-odd preconditioned equation
on the blocks which must be solved when calculating the block 
forces and the two equations on the full lattice
which must be solved when calculating the block-interaction force.
The corresponding numbers of saved solutions will be denoted by 
$p_1$, $p_2$ and $\tilde{p}_2$, respectively. On the blocks 
the extrapolation method is taken to be the ``minimal residual extrapolation''
recommended by Brower et al.~[\ref{BrowerEtAl}], while for
the solutions on the full lattice a polynomial extrapolation is used.

In the case of the block interaction force, 
the combination of the deflation acceleration and the 
chronological inversion method 
leads to the peculiar behaviour seen in the initial period of the
iteration number history shown in fig.~1.
What happens there is that the solution forecast 
rapidly improves in the first few steps and more than compensates
the loss of efficiency of the deflation subspace.
The accuracy of the solution forecast then saturates and the
solver iteration numbers start to grow until the subspace
is updated for the first time.

\subsection 3.3 Reversibility of the molecular-dynamics evolution

In order to guarantee the correctness of 
the HMC algorithm, the approximate integration of the 
molecular-dynamics equations must be reversible.
The multiple-time integration scheme used here 
fulfils this requirement, although in practice the reversibility can be
compromised by rounding errors and the chosen solver
tolerances.

When the deflation acceleration is switched on, 
the situation is complicated by the fact that the propagation of the deflation 
subspace is not reversible. The propagation of the solutions of 
the Dirac equation is also not reversible, but as both methods 
only serve to speed up the computation of the quark forces, 
the reversibility violations
caused by them are proportional to the solver tolerances [\ref{BrowerEtAl}].
These must therefore be chosen so that the reversibility is 
guaranteed to high precision (see subsect.~4.3).

\section 4. Tests of the accelerated algorithm

In this section the DD-HMC and the accelerated DD-HMC algorithm
are submitted to a speed test.
The HMC parameters are set to the same values in the two
cases so that any observed speed-up factors can be unambiguously attributed
to the deflation acceleration and
the chronological inversion method\kern1.5pt%
\footnote{$\dagger$}{\footnotefont%
An updated version of the DD-HMC code
is available under the terms of the GNU Public License (GPL) 
at {http://cern.ch/luscher/DD-HMC}.
}.

\subsection 4.1 Lattice parameters \& field ensembles

All tests reported below
were performed on a $64\times32^3$ lattice at inverse bare
coupling $\beta=5.3$, 
coefficient $\csw=1.90952$ [\ref{NPimp}] 
of the Sheikholeslami--Wohlert improvement term
and four values of the sea-quark 
hopping parameter $\ksea$ (see table~1).
At these points in parameter space,
the lattice spacing in physical units
is estimated to be $0.0784(10)$ fm,
while the ``pion'' mass $\Mpi$ ranges from about $618$ 
to $282$ MeV as $\ksea$ increases from $0.13590$ to
$0.13635$ 
[\ref{QCDlite}].

In order to obtain average timings with statistical errors
of at most a few percent, it suffices to perform a series of short simulations,
starting from a set of statistically independent representative
gauge-field configurations.
Ensembles of such field configurations were generated on the 
specified lattice as part of another project
by the authors of ref.~[\ref{QCDlite}] and 
were made available for the tests conducted here.

\subsection 4.2 HMC parameters

Following previous work [\ref{QCDlite}], the DD-HMC algorithm is set up on a
division of the lattice into blocks of size $8^4$. The length
$\tau$ of the molecular-dynamics trajectories and the integration
step numbers $N_0$, $N_1$ and $N_2$ are then set to the values 
quoted in table~1. With these choices,
trajectory acceptance rates of $80-85\%$ are achieved.
Recall that the step numbers refer to the gauge, block and 
block-interaction forces, respectively [\ref{SchwarzIII}]. In particular, 
the latter is evaluated at molecular-dynamics 
times separated by steps of size $\epstwo=\tau/N_2$.

The simulations performed at the lightest quark mass considered
($\ksea=0.13635$) are close to the edge of the stability range of 
the HMC algorithm [\ref{Stability}]. Following a suggestion of 
the PACS--CS collaboration [\ref{Kuramashi},\ref{Ukita}], the trajectory length
has been set to half the usual value in this case in order 
to reduce the rate of trajectories with large energy deficits 
at the end of the molecular-dynamics evolution.
The experience so far is
that this choice does not lead to larger autocorrelation
times (if measured in units of molecular-dynamics time) so that
the efficiency of the simulation remains practically the same.

\topinsert
%Blanke Zahl
\newdimen\digitwidth
\setbox0=\hbox{\rm 0}
\digitwidth=\wd0
\catcode`@=\active
\def@{\kern\digitwidth}
\tablecaption{Parameter values used in the test runs}
\vskip2.0ex
$$\vbox{\settabs\+&%
                  xxxxxxxxxx&x&%             Kappa
                  xxxxxxx&x&%                 tau
                  xxxxxxxxxxxxxx&x&%             N0,N1,N2
                  xxxxxxxxxxxxxx&x&\cr%          p1,p2,p2t
\thicktablerule
\vskip1.0ex
                \+& \hfill $\ksea$\hfill
                 && \hfill $\tau$\hfill
                 && \hfill $N_0$\kern0.7em$N_1$\kern0.7em$N_2$\hfill
                 && \hfill $p_1$\kern0.7em$p_2$\kern0.7em$\tilde{p}_2$\hfill
                 &\cr
\vskip1.0ex
\thintablerule
\vskip1.2ex
  \+& \hfill $0.13590$\hfill
  &&  \hfill $0.50$\hfill
  &&  \hfill $4$\kern1.2em$5$\kern1.2em$11$\kern1.3em
  &&  \hfill $8$\kern1.2em$6$\kern1.2em$5$\kern1.8em&\cr
\vskip0.3ex
  \+& \hfill $0.13610$\hfill
  &&  \hfill $0.50$\hfill
  &&  \hfill $4$\kern1.2em$5$\kern1.2em$14$\kern1.3em
  &&  \hfill $8$\kern1.2em$6$\kern1.2em$5$\kern1.8em&\cr
\vskip0.3ex
  \+& \hfill $0.13625$\hfill
  &&  \hfill $0.50$\hfill
  &&  \hfill $4$\kern1.2em$5$\kern1.2em$20$\kern1.3em
  &&  \hfill $8$\kern1.2em$6$\kern1.2em$5$\kern1.8em&\cr
\vskip0.3ex
  \+& \hfill $0.13635$\hfill
  &&  \hfill $0.25$\hfill
  &&  \hfill $4$\kern1.2em$5$\kern1.2em$16$\kern1.3em
  &&  \hfill $8$\kern1.2em$4$\kern1.2em$3$\kern1.8em&\cr
\vskip1.2ex
\thicktablerule
}
$$
\vskip-2.0ex
\endinsert

\subsection 4.3 Choice of the solver tolerances

The computation of the pseudo-fermion action and the 
fermion forces requires the Dirac equation
$D\psi=\eta$ (or the associated normal equation) to be solved 
numerically. In each case the algorithm used for this task
is stopped as soon as the calculated solution satisfies
$\|\eta-D\psi\|\leq \omega\|\eta\|$ for a specified tolerance
$\omega$.

For lattices of size $32\times24^3$, the tolerances recommended
in ref.~[\ref{SchwarzII}] were $10^{-8}$ for the block forces,
$10^{-7}$ for the block-interaction force and $10^{-11}$ and
$10^{-10}$ for the corresponding pseudo-fermion actions.
The same tolerances were also used in the simulations
reported in ref.~[\ref{QCDlite}], some of which were performed
on $64\times32^3$ lattices.

However, 
as already mentioned, the deflation acceleration and the 
chronological inversion method tend to compromise the reversibility
of the numerical integration of the molecular-dynamics equations. 
Extensive reversibility tests 
on the $64\times32^3$ lattice at $\ksea=0.13625$ actually show
that the tolerances should better
be set to $10^{-10}$ for the forces and $10^{-11}$ for the 
pseudo-fermion actions if the accelerations are switched on.
After a return trajectory, the absolute value of the energy deficit 
is then always less than $10^{-5}$ and the 
components of the initial and final gauge-field configurations
deviate by at most $10^{-9}$. 

Similar reversibility violations were observed
without acceleration and the previously recommended 
values of the tolerances.
In the speed tests, the tolerances were therefore taken to 
be the old ones in the case of the ordinary DD-HMC simulations
and the smaller ones specified above in the case of the 
accelerated simulations.

\subsection 4.4 Deflation-subspace and other parameters

At all values of the sea-quark mass, the
deflation subspace was generated by applying $9$
inverse iteration steps to $\Nm=20$ random quark fields and by
projecting them to a division of the lattice into 
blocks of size $4^4$.
This choice of parameters practically coincides with the
one suggested in ref.~[\ref{DDDeflation}].
Some experimenting was however required in order to find the
best values of the numbers $p_1$, $p_2$ and $\tilde{p}_2$ of 
old fields that are to be used
for the chronological propagation of the solutions of the Dirac equation 
(see table~1). 

The parameters of the Schwarz preconditioned GCR solver were 
set to the values previously recommended in 
refs.~[\ref{SchwarzII},\ref{DDDeflation}].
In particular, the size of the blocks on which the Schwarz preconditioner
operates was taken to be $8\times4^3$ in all cases.

\subsection 4.5 Test results

The figures listed in table~2 show that 
the average solver iteration numbers $N_{\rm GCR}$ 
required for the computation of the block-interaction force
are significantly reduced when the deflation acceleration
and the chronological inversion method are
switched on.
Most of the reduction is achieved through the 
deflation of the Dirac equation, 
the additional reduction through the 
solution forecast being at the level of 10--20\%.

\topinsert
%Blanke Zahl
\newdimen\digitwidth
\setbox0=\hbox{\rm 0}
\digitwidth=\wd0
\catcode`@=\active
\def@{\kern\digitwidth}
\tablecaption{Average solver iteration numbers $N_{\rm X}$
and execution times $t$ per trajectory} 
\vskip-1.0ex
$$\vbox{\settabs\+&%
                  xxxxxxxxxx&&%                         Kappa
                  xxxxxxxxxxxxxxxxxxxxxxxxxxxx&&%        Deflation off
                  xxxxxxxxxxxxxxxxxxxxxxxxxxxx&&\cr%     Deflation on
\thicktablerule
\vskip1.5ex
                \+&
                 && \hfill\hbox{\kern1.2em}DD-HMC\hfill
                 && \hfill\hbox{\kern1.2em}Accelerated DD-HMC\hfill
                 &\cr
\vskip0.4ex
                \+& \hfill $\ksea$\hfill
                 && \hfill $N_{\rm CG}$\kern1.5em$N_{\rm GCR}$%
                    \kern1.0em$t$ [min]\kern1.5em
                 && \hfill $N_{\rm CG}$\kern1.5em$N_{\rm GCR}$%
                    \kern1.0em$t$ [min]\kern1.5em
                 &\cr
\vskip1.0ex
\thintablerule
\vskip1.2ex
  \+& \hfill$0.13590$\hfill
  &&  \hfill $124$\kern0.9cm$@41$\kern2.4em$22$\kern2.3em
  &&  \hfill $ 59$\kern0.9cm$@17$\kern2.4em$17$\kern2.3em&\cr
\vskip0.3ex
  \+& \hfill$0.13610$\hfill
  &&  \hfill $126$\kern0.9cm$@60$\kern2.4em$36$\kern2.3em
  &&  \hfill $ 53$\kern0.9cm$@18$\kern2.4em$22$\kern2.3em&\cr
\vskip0.3ex
  \+& \hfill$0.13625$\hfill
  &&  \hfill $127$\kern0.9cm$@96$\kern2.4em$75$\kern2.3em
  &&  \hfill $ 48$\kern0.9cm$@20$\kern2.4em$32$\kern2.3em&\cr
\vskip0.3ex
  \+& \hfill$0.13635$\hfill
  &&  \hfill $128$\kern0.9cm$158$\kern2.4em$93$\kern2.3em
  &&  \hfill $ 40$\kern0.9cm$@22$\kern2.4em$28$\kern2.3em&\cr
\vskip1.2ex
\thicktablerule
}
$$
\vskip-2.0ex
\endinsert

Also shown in the table are the average
conjugate-gradient iteration numbers $N_{\rm CG}$
needed to compute the block forces. Here the 
observed reduction in the iteration numbers is a consequence of the 
solution forecast alone. Note that a reduction by a factor 2--3
is achieved even though the solver tolerance had to be lowered
in order to preserve the reversibility 
of the integration of the molecular-dynamics equations.

The average execution times $t$ per trajectory quoted in table~2 were
measured on a recent PC cluster with $32$ (single-core) dual-processor
nodes connected through a switched Infiniband network.
While these figures depend on program and hardware details,
they clearly show, in a realistic case,
that the algorithmic accelerations do result in important speed-up
factors.
They also lead to a softer scaling
of the timings with respect to the sea-quark mass (see fig.~2), 
as was to be expected given
the flat scaling behaviour of the deflated 
Schwarz-preconditioned GCR solver [\ref{DDDeflation}].

\topinsert
\vbox{
\vskip0.0cm
\epsfxsize=9.0cm\hskip1.25cm\epsfbox{plots/perf.eps}
\vskip0.3cm
\figurecaption{%
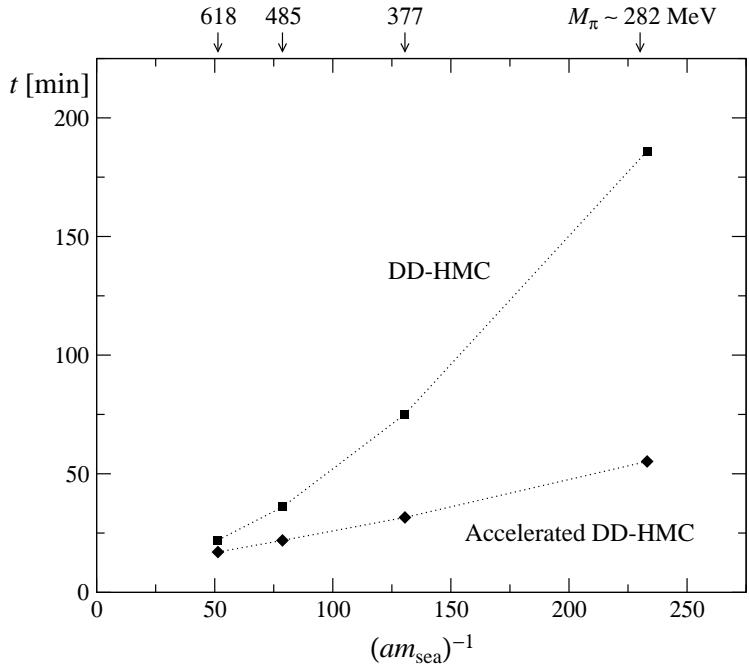
Plot of the average execution times $t$ (data points) 
required per molecular-dynamics trajectory
of length $\tau=0.5$ as a function of the inverse of the 
bare current-quark mass $\msea$ given in lattice units. 
At the smallest quark mass considered, 
the time needed for two trajectories 
of length $\tau=0.25$ is plotted. The lattice and algorithmic parameters
are specified in subsects.~4.1--4.4 and timings were taken on 
a PC cluster with $64$ processing units.
}
\vskip-0.0cm
}
\endinsert

The deflation acceleration tends to reduce the computer time
needed for the calculation of the block-interaction forces
to a level where the time spent in other parts of the program 
is not completely negligible anymore.
In the test runs, the
generation and propagation of the deflation subspace
consumed some 4--5 min per trajectory, while the bulk of the time
was divided roughly like 2:1 among the subprogram that computes
the block-interaction force and the remaining subprograms.

\section 5. Concluding remarks

Domain-decomposed deflation subspaces are technically attractive 
for many reasons. One of them certainly is the fact that
high deflation efficiencies can be achieved on large lattices
using fairly low numbers of modes per domain.
For the acceleration of the HMC algorithm,
another very important property is that these subspaces can be 
obtained with a modest computational effort.
Moreover, little extra work is required
to maintain their efficiency along the molecular-dynamics 
trajectories generated by the HMC algorithm.

The speed and excellent
scaling behaviour of the accelerated DD-HMC algorithm
are encouraging and make simulations
of lattice QCD with light Wilson quarks 
more feasible than ever before.
Deflation acceleration is, however, not
limited to the DD-HMC algorithm nor does one
have to follow the lines of this paper
in all detail.
In particular,
instead of the Schwarz alternating procedure, 
other relaxation algorithms can conceivably 
be used, both as preconditioner for the deflated GCR solver
and for the generation of the domain-decomposed deflation
subspaces.

\vskip1.0ex
I am indebted to Peter Weisz for his critical
comments on a first draft of the paper. 
Thanks also go to Bj\"orn Leder and Rainer Sommer for 
contributing machine-specific accelerations for IBM Blue Gene/L 
computers to the generic DD-HMC code 
and to Bj\"orn and Carlos Pena for their help in
validating the new version of the program. 
The gauge-field configurations
used for the numerical studies
were generated by the authors of ref.~[\ref{QCDlite}].
All computations reported here 
were performed on a dedicated PC cluster at CERN.

\appendix A. Subspace generation and update procedures

As explained in the main text, the domain-decomposed deflation subspaces
are generated and updated using approximate inverse iteration. 
In this appendix, the exact procedures that were used
in the test runs are described in some greater detail.

\subsection A.1 Field initialization and first steps

The inverse iteration (2.2) operates on the quark fields 
$\psi_l(x)$, $l=1,\ldots,\Nm$, that define
the domain-decomposed deflation subspace via the block projection (2.1).
At the beginning of each molecular-dynamics trajectory,
the components of these fields are initialized to 
uniformly distributed random values in the range $[-1,1]$.
The fields are then updated three times according to 
\equation{
   \psi_l(x)\to M_{\rm sap}\psi_l(x), 
   \enum
}
where $M_{\rm sap}$ denotes
the multiplicative Schwarz preconditioner
that was introduced in ref.~[\ref{SchwarzII}]. $M_{\rm sap}$ depends 
on several adjustable parameters, but their choice is not critical and
good results are obtained using similar
parameter values as in the case
of the Schwarz-preconditioned GCR solver.
However, as already mentioned in sect.~2, the
bare quark mass should be set to a value close to 
(or even equal to) the critical mass in this calculation,
independently of the value of the sea-quark mass.

\subsection A.2 Subspace refinement

After the initial phase of the subspace generation, the fields
$\psi_l(x)$ are updated several more times
using a deflated variant of approximate inverse iteration. 
A deflated inverse iteration step begins by
constructing the domain-decomposed deflation subspace
from the current set of fields through 
the block projection (2.1). This space is left unchanged
until all fields are updated once according to the rule
\equation{
   \psi_l(x)\to M_{\rm sap}\{\psi_l(x)-D\zeta_l(x)\}+\zeta_l(x),
   \enum
}
where $\zeta_l(x)=P\zeta_l(x)$ is an approximate solution of the equation
\equation{
   PDP\zeta_l(x)=\psi_l(x)
   \enum
}
in which $P$ denotes the 
orthonormal projector to the deflation subspace.

Note that the fields $\psi_l(x)$ are
contained in the deflation subspace by construction.
Equation (A.3) is thus a well-defined, square
linear system that actually coincides with 
the ``little Dirac equation'' of ref.~[\ref{DDDeflation}].
In appendix A of that paper, a preconditioned GCR solver 
for the little equation is described which is also used here.
Since inverse iteration is anyway
approximately implemented, there is no point in solving the
equation very accurately. In the test runs reported
in this paper, for example, the solver tolerance was set
to $10^{-3}$.

The total number of inverse iteration steps that must be applied
to generate an effective deflation subspace depends on the 
lattice parameters, the chosen block size and the number $\Nm$ of 
deflation modes per block. In the cases considered so far,
good deflation subspaces were obtained after
9--11 steps (3 ordinary plus 6--8 
deflated inverse iteration steps).

\subsection A.3 Subspace updates

As explained in sect.~3, the deflation subspace is updated
along the molecular-dy\-na\-mics trajectories by applying
a deflated inverse iteration step from time to time.
The procedure is the same as the one described above,
i.e.~the quark fields $\psi_l(x)$, $l=1,\ldots,\Nm$, that define
the current subspace are updated according to eq.~(A.2).

However, 
it is recommended to orthonormalize the fields
before they are updated,
as otherwise it can happen that they become more and more aligned
to each other as one moves along the trajectory.
After many steps, inverse iteration actually 
projects any field to the few very lowest modes of the Dirac operator, 
and although the gauge field changes along a trajectory,
the subspace updates can have a similar effect.

\appendix B. Automatising the subspace updates

The basic idea of the method proposed here is
to observe the solver iteration numbers along the trajectories 
(as in fig.~1) and to update
the subspace when the numbers have grown 
beyond a certain level since the last update.

Let $n(t)$ be the sum of the GCR solver iteration numbers
at molecular-dynamics time $t$ and suppose that the last
subspace update was at time $t_0$. In the 
following steps along the trajectory,
$n(t)$ will normally increase monotonically 
up to some time $t_1$, where the sum of the iteration number
differences $n(t)-n(t_0)$ satisfies the bound
\equation{
  \sum_{t=t_0}^{t_1}\{n(t)-n(t_0)\}\geq\frac{1}{2}\Nm.
  \enum
}
When this point is reached, the subspace is updated at the next step
of the molecular-dynamics integration and the procedure 
then repeats itself to the end of the trajectory.

The inequality (B.1) balances the effort required for the update of
the subspace against the additional work that is required for the
solution of the Dirac equation with respect to what it would be if the
deflation subspace were always in good condition. 
However, when the deflation acceleration
is combined with the chronological inversion method,
the left-hand side of the inequality should be modified so as to 
take into account the fact that the 
iteration numbers may not increase monotonically between
the subspace updates. A simple possibility is to replace
the differences $n(t)-n(t_0)$ in eq.~(B.1) by
\equation{
  n(t)-\min_{t_0\leq s\leq t}n(s)+
  \cases{
  {1\over16}n(t) & if $t<\max\{p_2,\tilde{p}_2\}\epstwo$,\cr
  \noalign{\vskip1ex}
              0           & otherwise,
  \cr}
  \enum
}
where $p_2$ and $\tilde{p}_2$
are the numbers of old fields used for the propagation of the
solutions of the first and the second Dirac equation 
on the full lattice (cf.~subsect.~3.2).

\beginbibliography

% Hasenbusch acceleration

\bibitem{Hasenbusch}
M. Hasenbusch,
Phys. Lett. B519 (2001) 177

\bibitem{HasenbuschJansen}
M. Hasenbusch, K. Jansen,
Nucl. Phys. B659 (2003) 299

\bibitem{DellaMorteEtAl}
M. Della Morte et al. (ALPHA collab.),
Comput. Phys. Commun. 156 (2003) 62

% Schwarz preconditioning

\bibitem{SchwarzI}
M. L\"uscher,
JHEP 0305 (2003) 052

\bibitem{SchwarzII}
M. L\"uscher,
Comput. Phys. Commun. 156 (2004) 209

\bibitem{SchwarzIII}
M. L\"uscher,
Comput. Phys. Commun. 165 (2005) 199

% Mass-preconditioned HMC

\bibitem{UrbachEtAl}
C. Urbach, K. Jansen, A. Shindler, U. Wenger,
Comput. Phys. Commun. 174 (2006) 87

% RHMC algorithm

\bibitem{RHMC}
M. A. Clark, A. D. Kennedy,
Phys. Rev. Lett. 98 (2007) 051601

% Hybrid Monte-Carlo algorithm for lattice QCD

\bibitem{HMC}
S. Duane, A. D. Kennedy, B. J. Pendleton, D. Roweth,
Phys. Lett. B195 (1987) 216

% Recent simulations using improved algorithms

\bibitem{QCDlite}
L. Del Debbio, L. Giusti, M. L\"uscher, R. Petronzio, N. Tantalo,
JHEP 0702 (2007) 056; {\it ibid.} 0702 (2007) 082

\bibitem{ETMC}
Ph. Boucaud et al. (ETM collab.),
Phys. Lett. B650 (2007) 304

\bibitem{FukayaEtAl}
H. Fukaya et al. (JLQCD and TWQCD collab.),
Phys. Rev. Lett. 98 (2007) 172001; Phys. Rev. D76 (2007) 054503

\bibitem{AlltonEtAl}
C. Allton et al. (RBC and UKQCD collab.),
Phys. Rev. D76 (2007) 014504

\bibitem{Kuramashi}
Y. Kuramashi, {\it Dynamical Wilson quark simulations toward the 
physical point}, plenary talk,
25th International Symposium on Lattice Field Theory,  
Regensburg 2007

\bibitem{Ukita}
N. Ukita et al. (PACS-CS collab.), 
in: Proceedings of the 
25th International Symposium on Lattice Field Theory,  
Regensburg 2007,
PoS(LATTICE 2007)138

% Low-mode deflation

\bibitem{deForcrand}
Ph. de Forcrand,
Nucl. Phys. B (Proc. Suppl.) 47 (1996) 228

\bibitem{NeffEtAl}
H. Neff, N. Eicker, T. Lippert, J. W. Negele, K. Schilling,
Phys. Rev. D64 (2001) 114509

\bibitem{MorganWilcox}
R. B. Morgan, W. Wilcox,
Nucl. Phys. (Proc. Suppl.) 106 (2002) 1067

\bibitem{LowModeKpipi}
L. Giusti, C. Hoelbling, M. L\"uscher, H. Wittig,
Comput. Phys. Commun. 153 (2003) 31

\bibitem{LowModeGiusti}
L. Giusti, P. Hern\'andez, M. Laine, P. Weisz, H. Wittig,
JHEP 0404 (2004) 013

\bibitem{LowModeDeGrand}
T. A. DeGrand, S. Schaefer,
Comput. Phys. Commun. 159 (2004) 185

\bibitem{BaliEtAl}
G. S. Bali, H. Neff, T. D\"ussel, T. Lippert, K. Schilling (SESAM collab.),
Phys. Rev. D71 (2005) 114513

\bibitem{AllToAllTrinity}
J. Foley, K. J. Juge, A. O'Cais, M. Peardon, S. M. Ryan, J.-I. Skullerud,
Comput. Phys. Commun. 172 (2005) 145

\bibitem{DDDeflation}
M. L\"uscher,
JHEP 0707 (2007) 081

% Chronological inversion method

\bibitem{BrowerEtAl}
R. C. Brower, T. Ivanenko, A. R. Levi, K. N. Orginos,
Nucl. Phys. B484 (1997) 353

% O(a) improved lattice QCD

\bibitem{SW}
B. Sheikholeslami, R. Wohlert,
Nucl. Phys. B259 (1985) 572

\bibitem{OaImp}
M. L\"uscher, S. Sint, R. Sommer, P. Weisz,
Nucl. Phys. B478 (1996) 365

% Determination of csw

\bibitem{NPimp}
K. Jansen, R. Sommer (ALPHA collab.),
Nucl. Phys. B530 (1998) 185
[E: {\it ibid.} B643 (2002) 517]

% Stability paper

\bibitem{Stability}
L. Del Debbio, L. Giusti, M. L\"uscher, R. Petronzio, N. Tantalo,
JHEP 0602 (2006) 011

% Multiple-time integration

\bibitem{SextonWeingarten}
J. C. Sexton, D. H. Weingarten,
Nucl. Phys. B380 (1992) 665

\endbibliography

\bye